\documentclass[fleqn,usenatbib]{mnras}

\usepackage{newtxtext,newtxmath}

\usepackage[T1]{fontenc}

\DeclareRobustCommand{\VAN}[3]{#2}
\let\VANthebibliography\thebibliography
\def\thebibliography{\DeclareRobustCommand{\VAN}[3]{##3}\VANthebibliography}

\usepackage{graphicx}	
\usepackage{amsmath}	

\RequirePackage{fix-cm}

\newcommand{\Msun}{M_{\odot}}
\newcommand{\Mstars}{M_{*}}
\newcommand{\BtoTkin}{\mathrm{B/T}}
\newcommand{\StarsBtoTkin}{\mathrm{(B/T)}_{*}}
\newcommand{\IdealStarsBtoTkin}{\mathrm{(B/T)_{*,ideal}}}

\newcommand{\ColdGasBtoTkin}{\mathrm{(B/T)_{cold \, gas}}}
\newcommand{\Rconv}{r_\mathrm{conv}}

\newcommand{\StarsSize}{r_{1/2, *}}

\newcommand{\Sersic}{S\'{e}rsic \,}

\title[Kinematic morphology of low-mass galaxies]{Kinematic morphology of low-mass galaxies in IllustrisTNG}

\author[Zeng et al.]{
Guangquan Zeng,$^{1,2}$\thanks{E-mail: zenggq@nao.cas.cn}
Lan Wang,$^{1,2}$\thanks{E-mail: wanglan@bao.ac.cn}
Liang Gao,$^{1,2,3,4}$
and Hang Yang$^{1,2}$
\\
$^{1}$National Astronomical Observatories, Chinese Academy of Sciences, Beijing 100101, China \\
$^{2}$School of Astronomy and Space Science, University of Chinese Academy of Sciences, Beijing 100049, China \\
$^{3}$School of Physics and Microelectronics, Zhengzhou University, Zhengzhou 450001, China\\
$^{4}$Institute for Frontiers in Astronomy and Astrophysics, Beijing Normal University, Beijing 102206, China \\
}

\date{Accepted XXX. Received YYY; in original form ZZZ}

\pubyear{2015}

\begin{document}
\label{firstpage}
\pagerange{\pageref{firstpage}--\pageref{lastpage}}
\maketitle

\begin{abstract}
The origin of diverse kinematic morphologies observed in low-mass galaxies is unclear.
In this study, we investigate the kinematic morphologies of central galaxies with stellar mass $10^{8.5-9.0}\Msun$ at $z=0$ in the TNG50-1 cosmological simulation.
The majority of the low-mass galaxies in TNG50-1 are dispersion-dominated, consistent with observations. 
By tracing the evolutionary histories of simulated low-mass galaxies, we find that while most stars form in rotating cold gas discs, the orientation of the star-forming discs relative to the galaxies may evolve with cosmic time.
If the cold gas disc remains aligning with the galaxy during its evolution, stars formed at different times share the same rotational direction, leading to a rotation-dominated system.
On the contrary, frequent misalignment of cold gas disc would result in a dispersion-dominated system.
In addition, we also find that the two-body scattering can have a non-negligible numerical heating effect on the simulated galaxy morphology, especially at central regions of galaxies and for relatively low-mass galaxies.
By comparing results of simulations with different resolutions, our results suggest that the simulated morphology of galaxies is roughly reliable when their number of stellar particles exceeds about $10^{4}$, and bulge morphology of galaxies can not be resolved robustly at the resolution level of TNG50-1.

\end{abstract}




\section{Introduction}
\label{sec:Intro}

Galaxies in the Universe have various morphologies, with the regular ones being classified into two main categories: disc galaxies and elliptical galaxies \citep[e.g.][]{1926ApJ....64..321H, 1959HDP....53..275D, 1961hag..book.....S, 1976ApJ...206..883V, 2012ApJS..198....2K, 2013MNRAS.435.2835W, 2015ApJS..217...32B, 2022MNRAS.509.3966W, 2023MNRAS.526.4768W}.
Disc galaxies have light distributions that can be fitted by an exponential profile, and display strong rotation kinematically, while in contrast, elliptical galaxies usually have an $n = 4$ \Sersic light profile \citep[][]{1963BAAA....6...41S} (i.e. the de Vaucouleurs profile) and are kinematically dispersion-dominated \citep[e.g.][]{1948AnAp...11..247D, 1999A&A...352..447C, 2011ApJS..196...11S, 2014MNRAS.439.1245K, 2015MNRAS.446.3943M, 2019MNRAS.483.2057F, 2024MNRAS.527..706Z}.

For low-mass galaxies (e.g. with stellar masses less than $10^{9} \Msun$), their morphology deviates from the general pattern of more massive galaxies.
Observations show that low-mass galaxies usually display a relatively irregular shape \citep[e.g.][]{1984AJ.....89..919S, 1984ARA&A..22...37G, 2014ARA&A..52..291C}.
Their light distributions can be fitted by a \Sersic profile with $n \lesssim 1$, similar to that of typical disc galaxies  \citep[e.g.][]{1983ApJ...266L..17F, 1994A&ARv...6...67F, 1998ARA&A..36..435M, 2012MNRAS.421.1007K, 2014MNRAS.439.1245K, 2015MNRAS.446.2967M, 2016ApJS..225...11Y, 2017MNRAS.468.4039R}.
On the other hand, kinematic measurements indicate that the majority of low-mass galaxies are dispersion-dominated, similar as the typical elliptical galaxies \citep[e.g.][]{2014ApJ...795L..37C, 2014MNRAS.439.1015K, 2015MNRAS.452..986S, 2017MNRAS.465.2420W, 2018NatAs...2..233Z, 2023ApJ...951...52D}.
Although evidences of rotation in stars and/or HI gas exist in some low-mass galaxies \citep[e.g.][]{2009A&A...493..871S, 2012AJ....144....4M, 2018MNRAS.477.1536E}.
For instance, \citet{2017MNRAS.465.2420W} analyzed 40 low-mass galaxies within the local volume and found that 32 of them are dispersion-dominated, while 8 are rotation-supported.

In hydrodynamical simulations, low-mass galaxies have kinematic morphologies qualitatively consistent with observations.
\citet{2018MNRAS.478.3994C}, \citet{2019MNRAS.487.5416T} and \citet{2018MNRAS.473.1930E} found that in EAGLE, TNG100-1 and FIRE simulations respectively, low-mass galaxies all generally exhibit dispersion-supported kinematics, with only a small fraction showing clear rotations.
Theoretically, galaxies are thought to firstly form during collapse of dark matter halos, and systems with larger angular momentum result in flatter morphology \citep[][]{1979Natur.281..200F, 1980MNRAS.193..189F, 1998MNRAS.295..319M}.
With subsequent mergers, galaxies may change their morphology and become elliptical ones \citep[][]{1977egsp.conf..401T, 1978MNRAS.183..341W, 1979Natur.281..200F}.
For low-mass galaxies having relatively quiet merger histories \citep[e.g.][]{2013MNRAS.428.3121M}, their morphologies are mostly affected by internal reasons \citep[e.g.][]{2007MNRAS.382.1187K, 2013pss6.book.....O, 2013MNRAS.429.3068T, 2013ARA&A..51..457N, 2014ARA&A..52..487S, 2017ARA&A..55...59N}.

What processes are responsible for the dispersion-dominated morphology of low-mass galaxies?
Some studies \citep[e.g.][]{2016ApJ...820..131E, 2017ApJ...835..193E, 2017MNRAS.465.1682H, 2017A&A...607A..13V, 2018MNRAS.473.1930E} pointed out that, due to the shallow gravitational potentials, the orderly motions of gas and stars in low-mass galaxies can be easily disrupted by various physical processes such as supernova explosions and stellar winds, making it difficult to form and maintain the rotation-dominated discs. 
Besides, some studies indicate that the gas compaction events may be necessary for the formation of rotational discs, which typically occur when galaxy mass reaches $\Mstars \approx 10^{9.5} \Msun$ \citep[e.g.][]{2014MNRAS.438.1870D, 2015MNRAS.450.2327Z, 2016MNRAS.457.2790T, 2016MNRAS.458..242T, 2016MNRAS.458.4477T, 2019MNRAS.488.4801J, 2020MNRAS.493.4126D, 2023MNRAS.522.4515L}.
Furthermore, \citet{2023MNRAS.525.2241H} showed that a centrally concentrated potential plays a crucial role in disc formation in simulated low-mass galaxies.
Nevertheless, it is still not clear why some low-mass galaxies, unlike most of the others, can still have disc shapes.

In addition to the physical factors, it should be noted that the numerical effects could also influence the kinematic morphology of simulated galaxies \citep[e.g.][]{2019MNRAS.488L.123L, 2020MNRAS.493.2926L, 2021MNRAS.508.5114L, 2023MNRAS.525.5614L, 2023MNRAS.519.5942W}.
The two-body scattering effect \citep[e.g.][]{1987gady.book.....B, 2008gady.book.....B} can cause energy exchange between stellar and dark matter particles in simulations, numerically resulting in a mass segregation of these two components, which can affect the size of simulated galaxies \citep[][]{2019MNRAS.488L.123L, 2020MNRAS.493.2926L}.
Moreover, the accumulation of two-body scattering can spuriously heat up an initially thin stellar disc, leading to a more spheroidal structure \citep[][]{2021MNRAS.508.5114L, 2023MNRAS.519.5942W}.
Therefore, when looking at morphology of simulated low-mass galaxies, the numerical heating effect needs to be treated with care.

In this work, using the zoom-in level cosmological simulation TNG50-1 \citep[][]{2019MNRAS.490.3234N, 2019MNRAS.490.3196P}, we study in more detail the origin and evolution of kinematic morphology of low-mass galaxies.
Our study aims to understand the basic reasons behind the diverse kinematic morphologies of low-mass galaxies, and explain why most low-mass galaxies are dispersion-dominated, rather than rotation-dominated.
Besides, we will assess the impact of numerical effect on simulated galaxy morphology, for galaxies with various stellar masses, and in simulations of different resolutions.

This paper is organized as follows.
In Section \ref{sec:Sample}, we introduce the simulations we use and how we select low-mass galaxies with different morphologies.
In Section \ref{sec:Results}, we investigate the kinematic morphology evolution and mass growth history of low-mass galaxies, and explore the reasons for their morphological differences.
In Section \ref{sec:Numerical}, we study the impact of numerical effect on the morphology of simulated galaxies.
Conclusions and discussions are presented in Section \ref{sec:Summary}.

\section{Simulation and Sample selection}
\label{sec:Sample}

\subsection{IllustrisTNG Simulations}
\label{subsec:TNG}
This work mainly use the publicly available data of TNG50-1 \citep[][]{2019MNRAS.490.3234N, 2019MNRAS.490.3196P}, which is the highest resolution realization of the IllustrisTNG project.
To investigate the impact of numerical resolution effect of the simulation on our results,  TNG50-2 and TNG50-3 are also investigated in \S \ref{subsec:numerical}.

The IllustrisTNG project is a series of cosmological hydrodynamical simulations of galaxy formation \citep[][]{2018MNRAS.475..624N, 2018MNRAS.480.5113M, 2018MNRAS.477.1206N, 2018MNRAS.475..676S, 2018MNRAS.475..648P}, built based on the fiducial TNG physics model \citep[][]{2018MNRAS.473.4077P, 2017MNRAS.465.3291W}.
The project consists of simulations with different volumes: TNG50, TNG100, and TNG300, which simulate galaxy formation within a periodic cubic box with a side length of 50, 100, and 300 Mpc, respectively.
For each volume, a series of realizations with different levels of resolution are conducted, labeled with suffixes `-1', `-2' or `-3' for decreasing resolutions.

In TNG50-1, the mass resolution of dark matter particle is $4.5 \times 10^5 \Msun$ and the baryonic mass resolution is $8.5 \times 10^4 \Msun$.

In IllustrisTNG simulations, the dark matter halos and subhalos are identified by with \texttt{FoF} and \texttt{Subfind} algorithms \citep[][]{2001MNRAS.328..726S, 2009MNRAS.399..497D} respectively, and the subhalos with stellar components are considered to be galaxies.
Merger trees of subhalos/galaxies are constructed by the \texttt{SubLink} algorithm \citep[][]{2015MNRAS.449...49R}.
In IllustrisTNG, cold-phase mass fraction of a star-forming gas cell is generally greater than $90 \%$ \citep[][]{2003MNRAS.339..289S}, therefore, following previous works \citep[e.g.][]{2018ApJS..238...33D, 2021MNRAS.507.3301Z, 2021MNRAS.507.4445N}, cold gas of a galaxy is represented by the gas with non-zero star formation rate.

\subsection{Selection of Low-Mass Galaxies}
\label{subsec:selection}

\begin{figure}
\centering
\includegraphics[width=0.9\columnwidth]{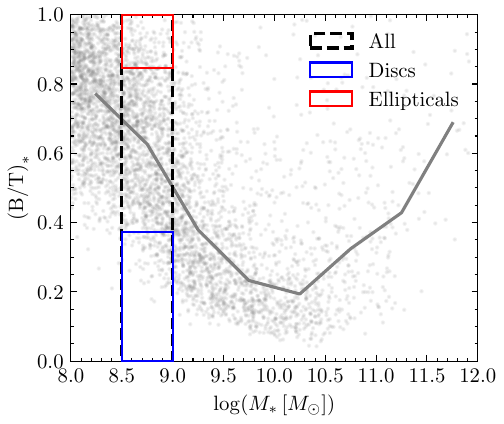}
\caption{
Stellar bulge-to-total mass ratio $\StarsBtoTkin$ as a function of galaxy stellar mass for central galaxies at $z=0$ in the TNG50-1 simulation.
Galaxies are denoted by gray dots, and the median relation is depicted by gray solid line.
The selected samples of low-mass galaxies, low-mass disc galaxies, and low-mass elliptical galaxies are included in the black, blue, and red boxed regions respectively.
Detailed criteria of our sample selection can be found in \S \ref{subsec:selection}.
}
\label{fig:stars_properties_z0_distribution}
\end{figure}

We select low-mass galaxies at $z=0$ in TNG50-1, with stellar mass $\Mstars$ ranging from $10^{8.5}$ to $10^{9.0} \Msun$.
In order to minimize the impact of environmental effects (e.g. tidal stripping) on the mass of such small galaxies, we only include galaxies that have consistently been central galaxies of their host halos since $z=6$. By doing so, 595 low-mass galaxies are selected.

Morphology of simulated galaxies are quantified using the kinematics-based bulge-to-total mass ratio \footnote{The galaxy properties such as mass, size, and morphology are all measured within $R_{200}$, unless specified otherwise.} \citep[e.g.][]{2009MNRAS.396..696S, 2012MNRAS.423.1726S}.
Firstly, the total angular momentum direction of all particles of interest (e.g. stars or gas cells) in the galaxy is set as $z$-axis, and the component of angular momentum of each particle along this axis is calculated as $j_z$.
Then the circularity parameter for each particle is defined as $\epsilon_{\mathrm{circ}} = j_z / j_\mathrm{circ}$, where $j_\mathrm{circ}$ refers to the specific angular momentum of a circular orbit at the radius of the particle.
The kinematics-based bulge-to-total mass ratio, $\BtoTkin$, is defined as twice the mass fraction of counter-rotating orbits with $\epsilon_{\mathrm{circ}} < 0$, and is used as the indicator of galaxy morphology.
$\BtoTkin =0$ corresponds to a galaxy with pure rotation, and $\BtoTkin =1$ corresponds to an ideal elliptical galaxy.

In Fig.~\ref{fig:stars_properties_z0_distribution}, gray dots represent the stellar bulge-to-total mass ratio $\StarsBtoTkin$ as a function of galaxy stellar mass, for all central galaxies at $z=0$ in TNG50-1.
Consistent with previous findings \citep[e.g.][]{2019MNRAS.487.5416T}, low- and high-mass simulated galaxies generally exhibit relatively large $\StarsBtoTkin$, while intermediate-mass galaxies with $\Mstars = 10^{9.5 \sim 10.5} \Msun$ tend to have a rotation-supported morphology.

We further select galaxies with the smallest $10 \%$ and largest $10 \%$ $\StarsBtoTkin$ as the disc and elliptical sub-samples, respectively, to study their different origins in the following analysis.
This gives 60 low-mass disc galaxies with $\StarsBtoTkin \lesssim 0.37$ and 60 low-mass elliptical galaxies with $\StarsBtoTkin \gtrsim 0.85$ from the total sample of 595 low-mass galaxies.
In Fig.~\ref{fig:stars_properties_z0_distribution}, the selected low-mass galaxies are marked out by boxed regions in various colors, with black dashed, blue solid, and red solid boxes including low-mass galaxies, low-mass disc galaxies, and low-mass elliptical galaxies, respectively.

\section{Origin of the morphology of low-mass galaxies}
\label{sec:Results}

In this section, we investigate why low-mass galaxies have diverse kinematic morphologies, and why most of them are dispersion-dominated.
We compare the morphological evolution and mass growth histories of low-mass galaxies with different morphologies, and check in detail the angular momentum directions of stars and cold gas at different times, to figure out the underlying reasons responsible for the observed morphological differences at the present-day.

\subsection{Morphology Evolution and Mass Growth History}
\label{subsec:MassEvolution}

\begin{figure}
\includegraphics[width=\columnwidth]{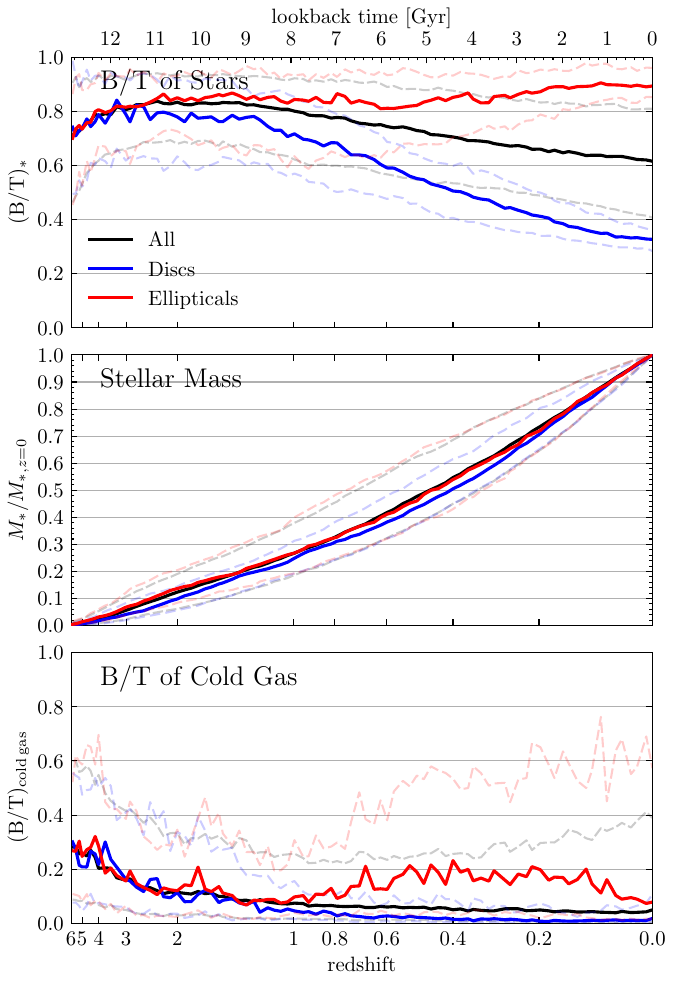}
\caption{
The evolution of the stellar bulge-to-total mass ratio $\StarsBtoTkin$ (top panel), galaxy stellar mass (middle panel, normalized by the present-day value), and the cold gas bulge-to-total mass ratio $\ColdGasBtoTkin$ (bottom panel) for all low-mass galaxies (black lines), low-mass disc galaxies (blue lines), and low-mass elliptical galaxies (red lines).
In each panel, solid lines represent the median relation and dashed lines include the 16th-84th percentile distributions.
}
\label{fig:diff_gals_morph_evolution}
\end{figure}

For the selected low-mass galaxies, we first check their morphological evolution with time, as presented in the top panel of Fig.~\ref{fig:diff_gals_morph_evolution}.
Black solid line indicates that, in general the stellar morphologies of low-mass galaxies tend to become more rotation-dominated starting from $z \sim 1$.
When selected by morphology, for low-mass disc galaxies, the $\StarsBtoTkin$ decreases significantly with time, and drops to smaller than $0.5$ after $z \sim 0.4$.
Low-mass elliptical galaxies, on the other hand, maintain a state of dispersion-dominated, with a roughly flat and even a slightly increasing of $\StarsBtoTkin$ after $z \sim 0.6$.
Overall, the morphological divergences between the two types of low-mass galaxies begin to emerge at $z \sim 2$, and widen up all the way to the present day.

In the middle panel of Fig.~\ref{fig:diff_gals_morph_evolution}, we present the stellar mass growth history of these low-mass galaxies.
Despite exhibiting different morphological evolution, these galaxies show very similar stellar mass evolution, with only disc galaxies having a bit later mass growth than others.
We have checked that for the low-mass galaxies, their stellar masses grow mostly in-situ, rather than from mergers with others galaxies, consistent with previous studies \citep[e.g.][]{2008MNRAS.384....2G, 2013MNRAS.428.3121M, 2016MNRAS.458.2371R, 2017MNRAS.467.3083R}.
In our low-mass galaxy sample, the mass fraction of stars formed in the central galaxy itself is mostly above $92 \%$, with a median value of $97.4 \%$.
For mergers these galaxies experience, the average number of mergers with a stellar mass ratio greater than $1:10$ is $0.40$ after $z=2$ and $0.16$ after $z=1$, with similar numbers across different galaxies.
Therefore, the morphological differences of these galaxies are not caused by different merger histories, and should come from the intrinsic properties of the galaxies themselves.

While stars form in cold gas component in a galaxy, we examine the kinematic property of cold gas in the low-mass galaxies, by looking at $\ColdGasBtoTkin$, which is defined similar as $\StarsBtoTkin$ but for cold gas particles as defined in \S \ref{subsec:TNG}.
As shown in the bottom panel of Fig.~\ref{fig:diff_gals_morph_evolution}, generally, most of the low-mass galaxies develop cold gas discs since early stages of evolution, consistent with the findings of \citet{2019MNRAS.490.3196P}.
The median trend of $\ColdGasBtoTkin$ for all low-mass galaxies decreases towards lower redshift and stays below $0.1$ after $z=2$.
Low-mass disc galaxies have a smaller median and a narrower scatter of $\ColdGasBtoTkin$, indicating highly rotational and stable cold gas discs.
Low-mass ellipticals show an obvious higher $\ColdGasBtoTkin$ with large scatter, having less rotational cold gas components. 

Note that most of the low-mass galaxies have $\ColdGasBtoTkin$ less than $0.3$ at all redshift, even for the present-day ellipticals.
In contrast, $\StarsBtoTkin$ of the stars is always much higher.
This indicates that while cold gas forms rotational discs and new stars are born within it, the star component in total somehow become more dispersion-dominated.
One possibility is that new stars born at each time do not maintain as much their original rotational kinematic state at later times.
We will demonstrate in Section \ref{sec:Numerical} that this could happen due to numerical heating effect.
The other possibility is that new stars born at different times have different rotation directions, and are mixed up thus resulting in a dispersion-dominated system, which is studied in the following subsection.

\subsection{Misaligned Star Formation}
\label{subsec:misaligned}

\begin{figure*}
\hspace{-0.4cm}
\resizebox{15cm}{!}{\includegraphics{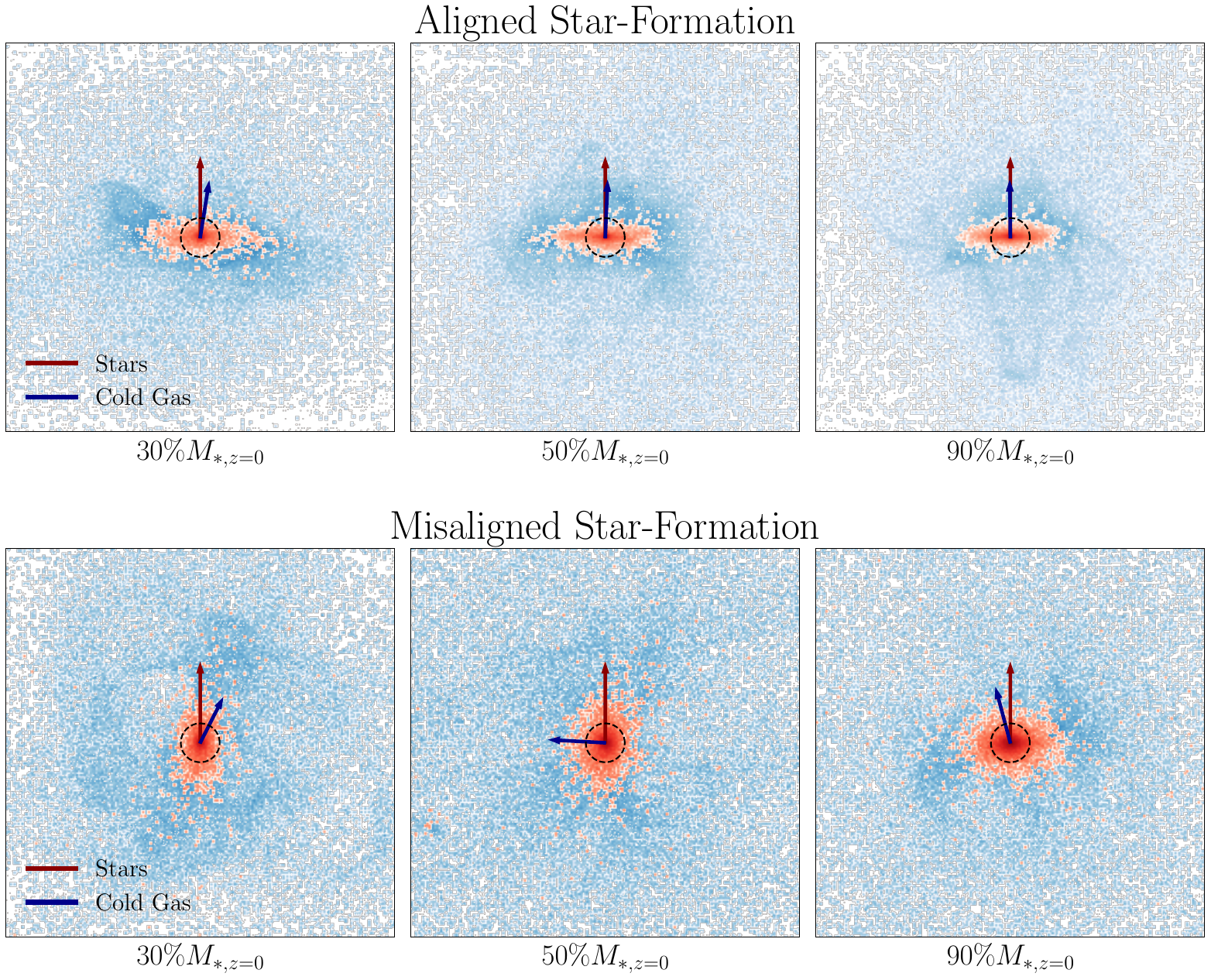}}
\caption{
Two examples of galaxies that have aligned star-formation (top row) and misaligned star-formation (bottom row).
For each example galaxy, panels from left to right show the density projections, with stars in red and gas in blue, at times when the galaxy has formed $30\%$, $50\%$, and $90\%$ of their stellar mass at $z=0$, respectively.
In each panel, the density projection is displayed from an edge-on view of the galaxy, and the angular momentum directions for stars and cold gas are indicated by the arrows. 
The dashed-line circle in each panel represents twice the half-stellar mass radius.
}
\label{fig:case_misaligned_star_formation}
\end{figure*}

\begin{figure}
\includegraphics[width=\columnwidth]{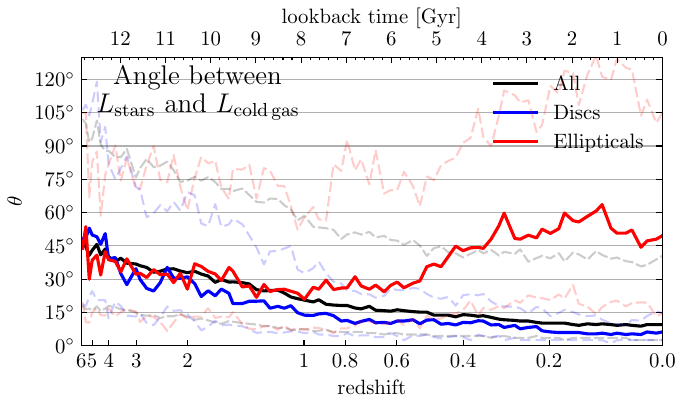}
\caption{
Evolution of the angle between the angular momentum direction of stars and that of cold gas for the low-mass galaxies.
Solid lines give the median relations and dashed lines include the 16th-84th percentile distributions.
}
\label{fig:diff_gals_theta_evolution}
\end{figure}

\begin{figure}
\centering
\includegraphics[width=0.85\columnwidth]{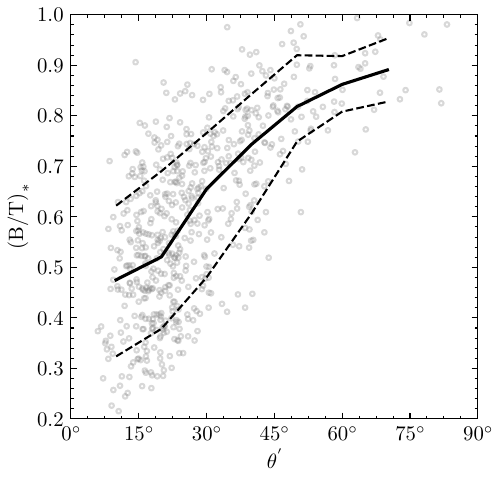}
\caption{
For all low-mass galaxies, relation between their stellar bulge-to-total ratio $\StarsBtoTkin$ and the mass-weighted misalignment angle ${\theta}^{'}$ (see \S \ref{subsec:misaligned} for the detailed definition).
Black solid line represents the median relation, and dash lines depict the 16th-84th percentile distribution.
It can be seen that, galaxies with larger ${\theta}^{'}$ (i.e. more misaligned star-formation) have more dispersion-dominated morphologies.
}
\label{fig:all_gals_misalignment_angle_vs_starsBtoTkin}
\end{figure}

To study why stars born in rotational discs of various times form much more dispersion-dominated systems over time, as shown in Fig.~\ref{fig:diff_gals_morph_evolution}, we check in detail how star-forming cold gas discs align with existing stellar components during the lifetimes of low-mass galaxies, and how much the alignment and mis-alignment affect the final stellar morphology of galaxies.

We look at and compare the directions of cold gas disc and stellar components, and find that the evolution of the alignment between the two can behave quite differently for different low-mass galaxies.
In Fig.~\ref{fig:case_misaligned_star_formation} we present two representative examples, corresponding to an aligned (top row) and a misaligned (bottom row) case, respectively.
For each galaxy, we show the rotation directions of the stars (red arrow) and the cold gas component (blue arrow), at three evolution stages in three panels.
As shown, for the galaxy shown in the top row, the angular momentum direction of the star-forming cold gas closely aligns with that of the stellar component at different stages.
On the contrary, for the galaxy presented in the bottom row, its cold gas component is relatively misaligned with the stars.
The direction of cold gas disc and hence the direction of the following newly-born stars change with time with respect to the existing stars.

Quantitatively, we measure the angle $\theta$ between the angular momentum of stars and that of cold gas at different times for the low-mass galaxies, and show the statistical result of the evolution of $\theta$ in Fig.~\ref{fig:diff_gals_theta_evolution}.
For all low-mass galaxies in general, $\theta$ decreases over time, implying an increasing of alignment between cold gas and existing stars with time. 
Low-mass discs have similar trend of $\theta$ evolution as the whole sample, with a lower $\theta$ value and stronger alignment at almost all redshifts.
Low-mass elliptical galaxies, on the other hand, have an obvious increase in $\theta$ starting from $z \sim 1$, reflecting their predominantly misaligned cold gas disc and star-formation.

For a given galaxy, to evaluate the overall extent of misaligned star-formation over its lifetime, we define a mass-weighted misalignment angle:
\begin{equation}
\label{eq:misaligned}
{\theta}^{'} = \Sigma \frac{M_{*,n} - M_{*,n-1}}{M_{*,z=0}} {\theta}_{n}
\end{equation}
where $M_{*,n}$ is the stellar mass at the $n$th snapshot for a given simulated galaxy, and ${\theta}_{n}$ is the angle between the angular momentum direction of cold gas and stars at that time.
A larger ${\theta}^{'}$ implies that the newly-formed stars in the galaxy during evolution are mostly misaligned with existing stars, whereas a smaller ${\theta}^{'}$ indicates that the galaxy mainly has an aligned cold gas component and hence star-formation through its lifetime.

In Fig.~\ref{fig:all_gals_misalignment_angle_vs_starsBtoTkin}, we present $\StarsBtoTkin$ for all low-mass galaxies as a function of their mass-weighted misalignment angle ${\theta}^{'}$.
It can be seen that, there exists a clear correlation between the misalignment angle and the final stellar morphology of these low-mass galaxies.
Specifically, low-mass galaxies with a larger ${\theta}^{'}$, experiencing misaligned star formation more frequently, tend to become more dispersion-dominated with a larger $\StarsBtoTkin$ at the present day.
And for those with a smaller ${\theta}^{'}$, the newly-formed stars during evolution are usually aligned with the overall galaxy, resulting in a more rotation-dominated morphology.
Therefore, Fig.~\ref{fig:all_gals_misalignment_angle_vs_starsBtoTkin} indicates that the extent of misalignment in star formation plays a crucial role in determining the final kinematic morphology of low-mass galaxies.
Frequent misalignment of star formation would lead to a dispersion-dominated morphology, while a consistently aligned star formation history would in general form a disc-like low-mass galaxy.
Similar mechanisms have been found in more massive galaxies in previous studies of numerical simulations \citep[e.g.][]{2012MNRAS.423.1544S, 2019ApJ...883...25P, 2022MNRAS.515..213P, 2022MNRAS.514.2031C}.

\section{Two-body Scattering heating on the simulated galaxy morphology}
\label{sec:Numerical}

As introduced in Section \ref{sec:Intro}, the morphology of simulated galaxies could be affected by the two-body scattering effect.
In this section, we investigate in detail the impact of this effect on low-mass simulated galaxies.
We also explore further how this numerical heating effect affects the morphology of galaxies with varying masses and simulated by varying resolutions.

\subsection{Two-Body Scattering Effect: Case Study}
\label{subsec:numerical_heating_case_study}

\begin{figure*}
\hspace{-0.4cm}
\resizebox{15cm}{!}{\includegraphics{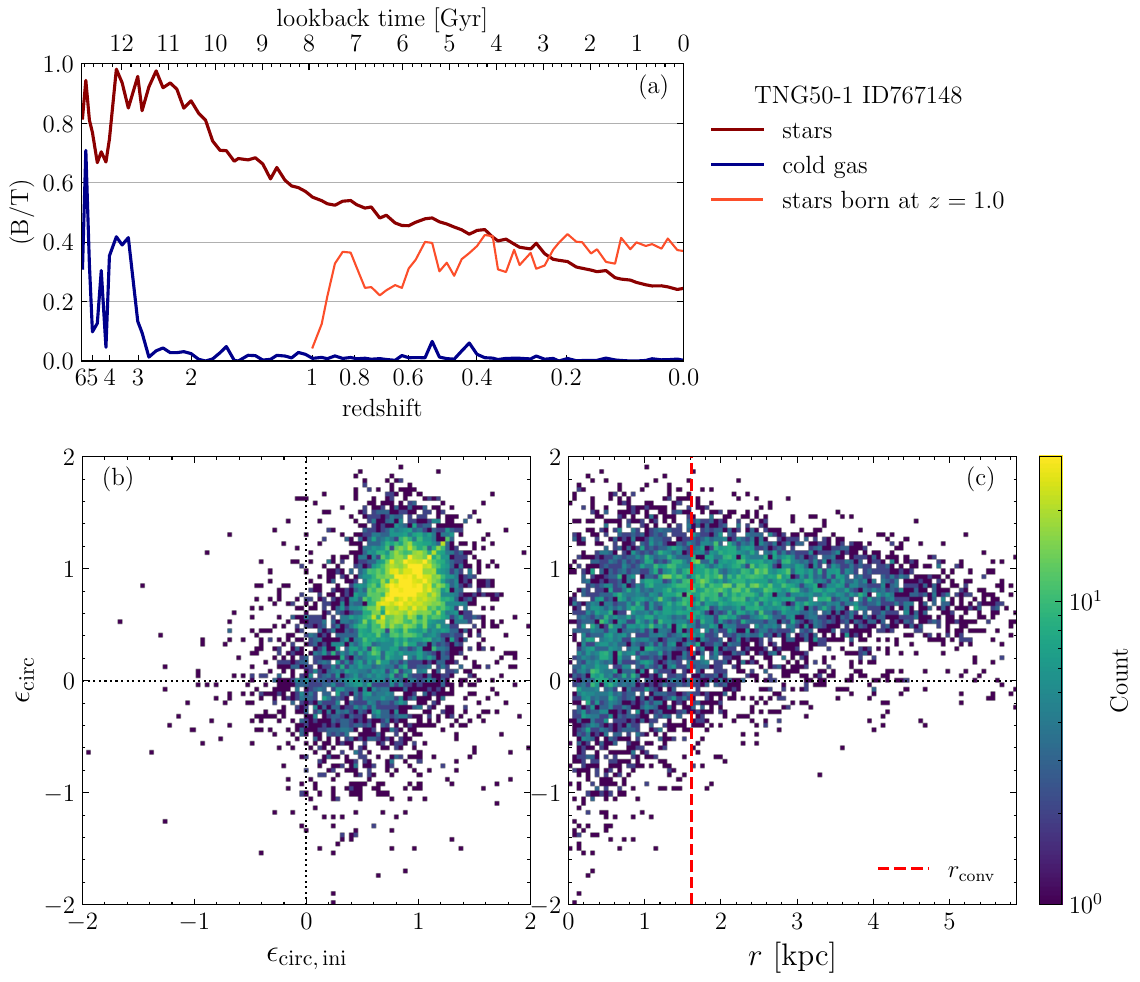}}
\caption{
A representative galaxy is presented to demonstrate the two-body scattering effect.
Panel (a) shows the evolution of $\StarsBtoTkin$ for the stellar component (red line), $\ColdGasBtoTkin$ for the cold gas component (blue line), and the evolution of $\StarsBtoTkin$ for the stars formed at $z=1$ (orange line).
For all stellar particles of this galaxy at the present day: panel (b) compares their present-day circularities, $\epsilon_{\mathrm{circ}}$ with their initial circularities at birth, $\epsilon_{\mathrm{circ,\, ini}}$; 
panel (c) gives the relation between their present-day circularities $\epsilon_{\mathrm{circ}}$ and their distances from the galactic center.
In panel (c), the red dashed line indicates the convergence radius $\Rconv$ of this galaxy (see text for details).
}
\label{fig:case_two_body_scattering}
\end{figure*}

Although a clear correlation between $\StarsBtoTkin$ and ${\theta}^{'}$ is demonstrated in Fig.~\ref{fig:all_gals_misalignment_angle_vs_starsBtoTkin}, it shows a considerable scatter, especially for low ${\theta}^{'}$.
For example, low-mass galaxies with ${\theta}^{'}$ less than $15^{\circ}$ have $\StarsBtoTkin$ ranging from approximately $0.2$ to $0.7$. 
This indicates that, despite having aligned star formation during evolution, the morphology of these galaxies are additionally influenced by other factors.
By examining the evolution of the stellar particles distributions of these galaxies further, we find that after born in cold rotating gas discs, some stars are subsequently heated to various degrees by the two-body scattering effect, and result in a kinematically hotter component than the cold gas disc.

In Fig.~\ref{fig:case_two_body_scattering}, we present a representative case which has a small mass-weighted misalignment angle ${\theta}^{'}$ of ${12}^{\circ}$, to illustrate the two-body scattering effect on simulated galaxies. 
For this example galaxy, panel (a) gives the evolution of $\StarsBtoTkin$ for the stellar component by red line, and $\ColdGasBtoTkin$ for the cold gas component by blue line.
Similar as already seen for most low-mass galaxies in Fig.~\ref{fig:diff_gals_morph_evolution}, this galaxy develops and maintains a fast rotating cold gas disc component since $z \sim 3$, while its stellar component is clearly much hotter.

In addition, the orange line in panel (a) shows the evolution of $\StarsBtoTkin$ for the stars formed within the preceding $0.1$ Gyr time interval of $z = 1$, to trace the kinematic evolution of stars after formation.
This population of stars formed at $z = 1$ is born cold within the gas disc and has a consistent kinematic state with the cold gas disc, having an initial $\StarsBtoTkin \sim 0.05$.
Then the stars increase their $\StarsBtoTkin$ dramatically in a relatively short period of time, and stay in a hotter state of $\StarsBtoTkin \sim 0.3-0.4$.
Similar phenomena are also found for stars born at other redshifts, and the increase in $\StarsBtoTkin$ can be smaller or larger.
In some cases, the kinematics of stellar particles can change significantly in just one snapshot after their formation, even without notable variations in the potential well.
This is mainly due to the two-body scattering between stellar particles and dark matter particles, as previously demonstrated by \citet{2021MNRAS.508.5114L} and \citet{2023MNRAS.519.5942W}

In the panel (b) of Fig.~\ref{fig:case_two_body_scattering}, for all stellar particles in this galaxy at present day, we further compare their present-day circularities $\epsilon_{\mathrm{circ}}$ (as defined in \S \ref{subsec:selection}) with their initial circularities $\epsilon_{\mathrm{circ,ini}}$ at time of birth.
Initially, the majority of the stellar particles have circularities concentrating around $1$ when newly formed, indicating rotating stars formed in the rotating cold gas discs.
However, at the present day, more stellar particles have decreasing circularities, and have evolved from co-rotating stars into counter-rotating stars, compared with the ones evolve oppositely.
In panel (c), we check the distribution of present-day circularities of stars as a function of radius.
It can be seen that the counter-rotating stars today with $\epsilon_{\mathrm{circ}} <0$ are mainly located in the inner regions, with a higher proportion toward the center.
Fig.~\ref{fig:case_two_body_scattering} demonstrate that, the kinematic morphology of simulated low-mass galaxies can be affected by the two-body scattering effect, more strongly in the central region.

\subsection{Quantifying Two-Body Scattering Effect}
\label{subsec:quantifying_numerical_heating}

\begin{figure}
\includegraphics[width=\columnwidth]{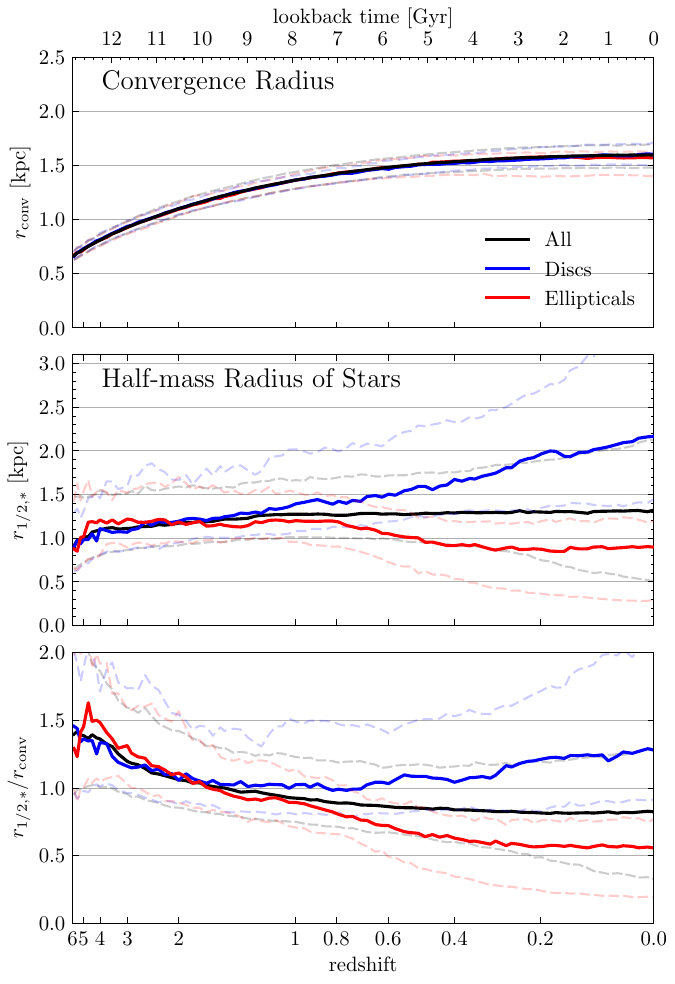}
\caption{
The evolution of convergence radius (top panel), stellar half-mass radius (middle panel), and the ratio of stellar half-mass radius to convergence radius (bottom panel) for all low-mass galaxies (black), low-mass disc galaxies (blue) and low-mass elliptical galaxies (red).
Solid line for each subsample represents the median evolution, and the region between the dashed lines indicates the 16th-84th percentile range.
}
\label{fig:diff_gals_hmr_evolution}
\end{figure}

\begin{figure}
\includegraphics[width=\columnwidth]{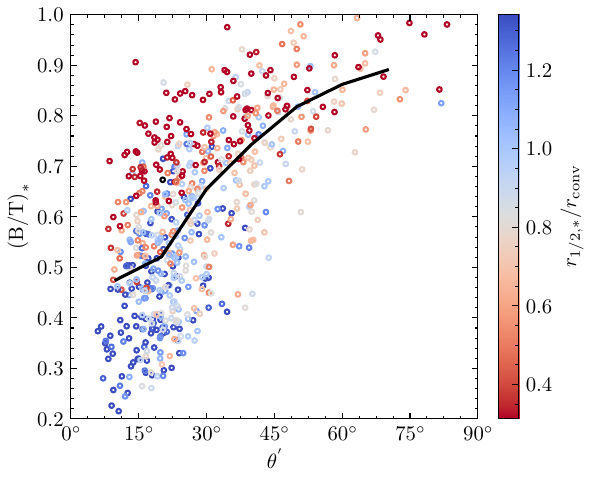}
\caption{
Similar as Fig.~\ref{fig:all_gals_misalignment_angle_vs_starsBtoTkin}, but each symbol is color-coded by $\StarsSize / \Rconv$ of the low-mass galaxies, with bluer symbols representing a larger size relative to the convergence radius.
At given misalignment angle, low-mass galaxies with larger relative sizes are less affected by the two-body scattering heating and have more rotation-dominated morphologies.
}
\label{fig:all_gals_misaligned_vs_hmr_vs_z0_stars_BtoTkin}
\end{figure}


The two-body scattering effect is commonly quantified by inspecting the two-body relaxation process theoretically \citep[e.g.][]{1987gady.book.....B, 2008gady.book.....B}, and also in simulations \citep[e.g.][]{2003MNRAS.338...14P, 2004MNRAS.348..977D,  2019MNRAS.487.1227Z, 2019MNRAS.488L.123L, 2019MNRAS.488.3663L}.
In this work, we assess the two-body relaxation for simulated galaxies by computing their corresponding convergence radius, $r_{\mathrm{conv}}$.
As discussed in \citet{2003MNRAS.338...14P} and \citet{2019MNRAS.488L.123L, 2019MNRAS.488.3663L}, beyond $r_{\mathrm{conv}}$, the time required for two-body relaxation is comparable to or greater than the Hubble time, and the impact of two-body scattering is considered to be small.
While within $r_{\mathrm{conv}}$, the two-body scattering effect can significantly affect the kinematics of simulated galaxies.

To calculate the convergence radius, we basically follow the method of \citet{2019MNRAS.488L.123L}, except with further consideration of the contribution from stellar particles.
For simulated galaxies, the two-body relaxation time $t_{\mathrm{relax}}$ is defined as the time required for significant velocity change of a particle due to cumulative two-body encounters with other particles.
Based on it, the convergence radius $r_{\mathrm{conv}}$ of simulated galaxies is estimated by requiring the two-body relaxation time at the convergence radius $t_{\mathrm{relax}, r_{\mathrm{conv}}}$ to be equal to the orbital time at the virial radius $t_{\mathrm{orbit}, 200}$, i.e., $t_{\mathrm{relax}, r_{\mathrm{conv}}} / t_{\mathrm{orbit}, 200} =1$.
More details on the calculations can be found in Appendix \ref{sec:Appendix_A}.

For the example galaxy shown in Fig.~\ref{fig:case_two_body_scattering}, its convergence radius is indicated by a red vertical dashed line in panel (c).
It can be seen that most of the counter-rotating stars in this galaxy are indeed distributed within $\Rconv$.
Beyond $\Rconv$, the initially-cold stellar disc of this galaxy remains well-preserved.
We examine all low-mass galaxies in our sample and find very similar behaviors.
Generally, these simulated galaxies are likely to experience less spurious heating from the two-body scattering effect if more stars are distributed beyond the convergence radius.

We examine the evolution of convergence radius for low-mass galaxies, to check whether their different morphologies are a result of different size of $\Rconv$.
In the top penal of Fig.~\ref{fig:diff_gals_hmr_evolution}, results show that these galaxies exhibit very similar evolution of the convergence radius, independent of their morphologies.
This implies that during evolution, these low-mass galaxies are consistently affected by two-body scattering within a similar radius range.

We then explore the evolution of the stellar half-mass radius $\StarsSize$ for different low-mass galaxies in the middle penal of Fig.~\ref{fig:diff_gals_hmr_evolution}.
It can be seen that low-mass disc galaxies have an obvious increase in size with redshift, while ellipticals have a decrease in size after redshift of around $0.8$.
This indicates that, despite having similar stellar mass and stellar mass evolution histories as shown in Fig.~\ref{fig:diff_gals_morph_evolution}, low-mass disc galaxies generally have a more extended distribution of stars compared to the ellipticals.
In the bottom panel of Fig.~\ref{fig:diff_gals_hmr_evolution}, the evolution of the ratio of stellar half-mass radius to convergence radius is presented.
For the low-mass ellipticals, the ratio shows a clear downward trend over time.
After $z\sim 2$, their sizes become comparable to the convergence radius and then evolve to only about a half at the present day, with most of their stellar component largely affected by two-body scattering heating.
Conversely, low-mass disc galaxies have more stars residing outside $\Rconv$ for most of the time during evolution, less affected by numerical heating, and therefore maintain their disc structure.

In Fig.~\ref{fig:all_gals_misaligned_vs_hmr_vs_z0_stars_BtoTkin}, we explore the impact of two-body scattering on the kinematic morphology for low-mass galaxies, in addition to the factor caused by misaligned star formation as studied in Fig.~\ref{fig:all_gals_misalignment_angle_vs_starsBtoTkin}.
This figure is similar as Fig.~\ref{fig:all_gals_misalignment_angle_vs_starsBtoTkin}, except that symbols that represent galaxies are color-coded by their ratios of stellar half-mass radius to convergence radius.
As shown, the large scatter of $\StarsBtoTkin$ at small misalignment star formation angle ${\theta}^{'}$ is mainly due to variations in $\StarsSize / \Rconv$, with galaxies having large relative sizes being more rotation-dominated.
Fig.~\ref{fig:all_gals_misaligned_vs_hmr_vs_z0_stars_BtoTkin} clearly indicates that in general two conditions are required to form the low-mass disc galaxies in simulations: first, a continuous process of aligned star-formation  (small ${\theta}^{'}$), and second, a galaxy size larger than the convergence radius (large $\StarsSize / \Rconv$) to minimize the numerical heating induced by the two-body scattering.

\subsection{Numerical Heating Effect on Galaxies of Various Masses and Resolutions}
\label{subsec:numerical}

\begin{figure*}
\hspace{-0.4cm}
\resizebox{13.5cm}{!}{\includegraphics{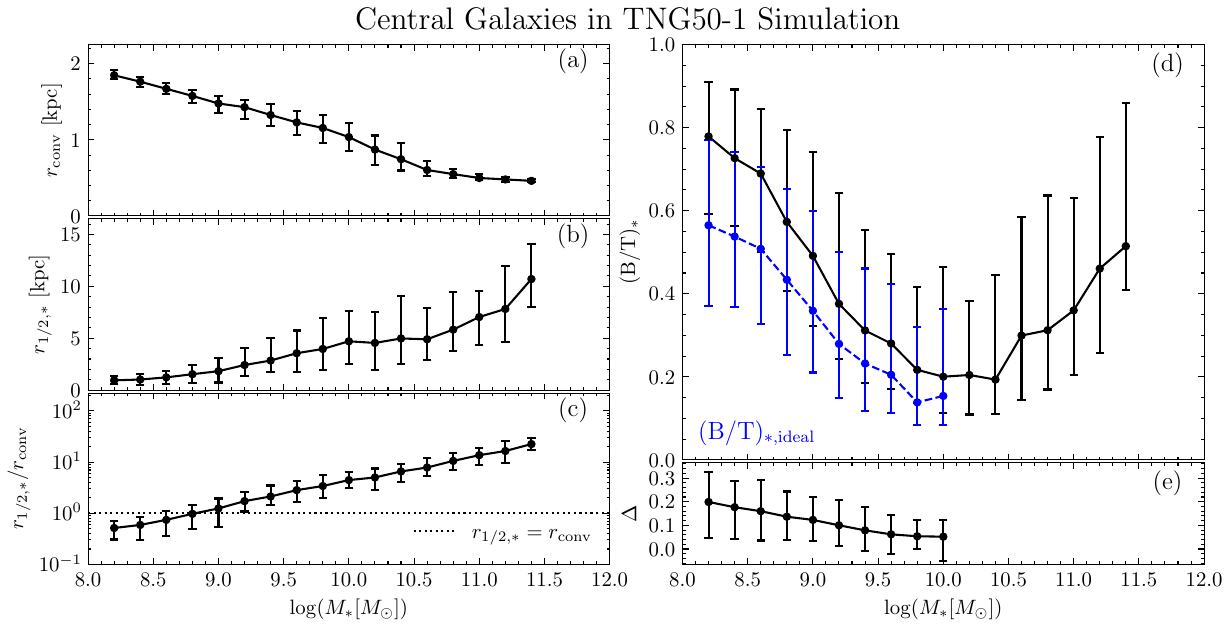}}
\caption{
For central galaxies at $z=0$ in TNG50-1, panels (a), (b), and (c) respectively present the convergence radius $\Rconv$, the stellar half-mass radius $\StarsSize$, and the ratio of stellar half-mass radius to convergence radius $\StarsSize / \Rconv$, as a function of galaxy stellar mass.
In addition, panel (d) shows the galaxy morphology $\StarsBtoTkin$ and ideal galaxy morphology $\IdealStarsBtoTkin$ (see text for the detail) as a function of stellar mass, and panel (e) illustrates the difference between the two, $\Delta = \StarsBtoTkin - \IdealStarsBtoTkin$, also as a function of stellar mass.
In each panel, the median trend is shown by black solid line, and error bars indicate the 16th-84th percentile distributions.
}
\label{fig:size_vs_morph}
\end{figure*}

\begin{figure*}
\hspace{-0.4cm}
\resizebox{17cm}{!}{\includegraphics{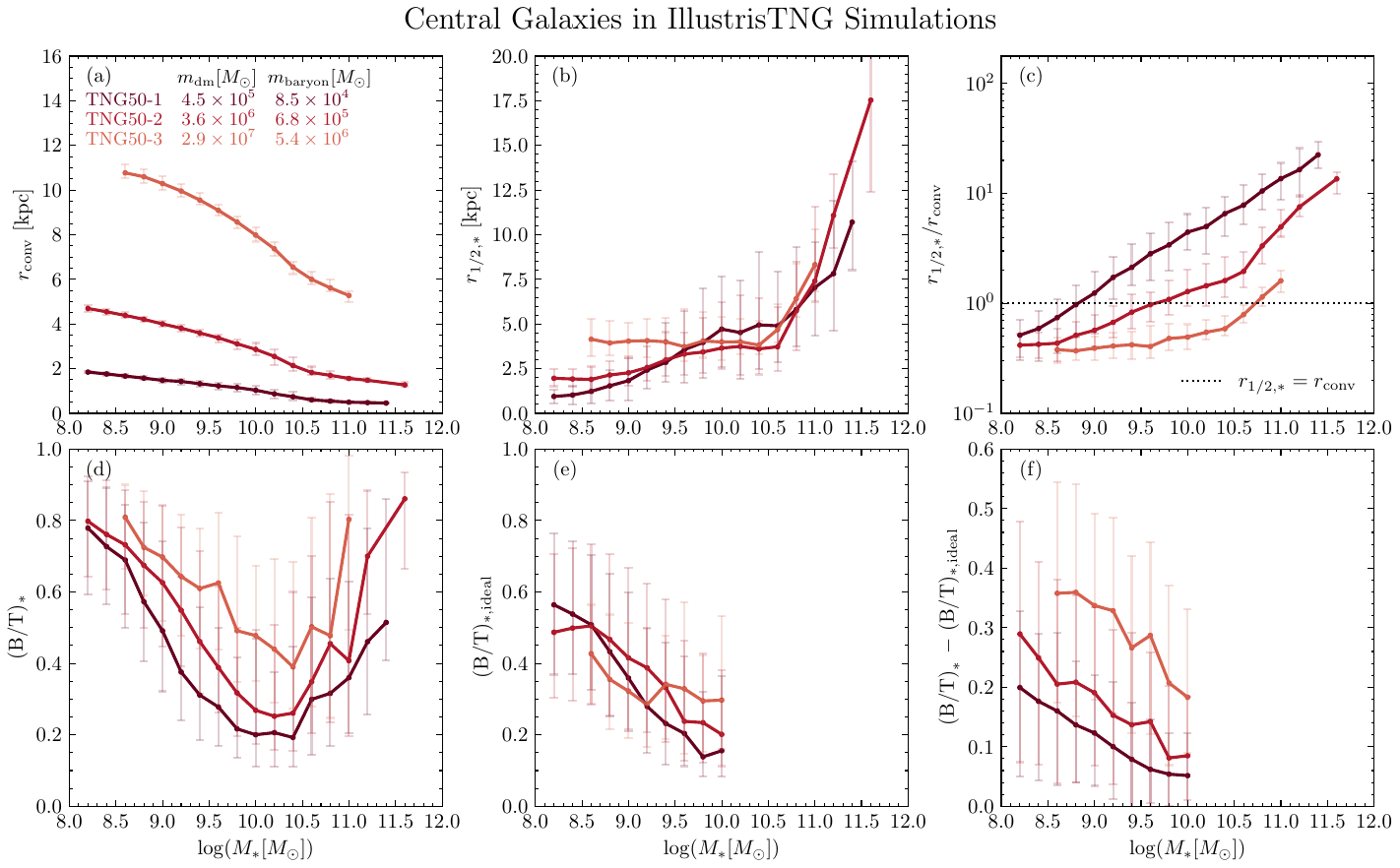}}
\caption{
Effect of resolution on simulated galaxy morphology is shown by comparing the results from three simulations of TNG50-1, TNG50-2 and TNG50-3.
Relations shown in this figure are similar to that in Fig.~\ref{fig:size_vs_morph}, with the properties of $\StarsBtoTkin$ and $\IdealStarsBtoTkin$ plotted separately in panels (d) and (e), and their difference plotted in panel (f), to see the results from these simulations more clearly.
Different simulations are represented using different colors, and the resolution of each simulation is indicated in the upper left corner of panel (a).
Central galaxies with a number of stellar particles greater than $100$ are studied, and statistics in mass bins with more than 10 galaxies are presented.
}
\label{fig:compare_diff_simulations}
\end{figure*}

In \S \ref{subsec:numerical_heating_case_study} and \S \ref{subsec:quantifying_numerical_heating}, we have shown the non-negligible role of two-body scattering effect in affecting the morphology of low-mass galaxies simulated in TNG50-1.
In this subsection, we further explore how this effect influences the simulated galaxies with different masses and resolutions.

To study the impact of two-body scattering on simulated galaxies with different masses, we select central galaxies at $z=0$ in TNG50-1 with stellar masses ranging from $10^{8}$ to $10^{12} \Msun$.
In Fig.~\ref{fig:size_vs_morph} we present their convergence radius $\Rconv$ (panel (a)), half-mass radius $\StarsSize$ (panel (b)) and $\StarsSize / \Rconv$ (panel (c)) as a function of stellar mass.
As galaxy stellar mass increases, the convergence radius decreases, and galaxy size increases.
As a result, $\StarsSize / \Rconv$ increases with stellar mass, and is greater than $1$ for galaxies more massive than $\sim 10^{9} \Msun$, with stars mainly distributed beyond the convergence radius and less affected by numerical heating.

In the panel (d) of Fig.~\ref{fig:size_vs_morph}, we plot $\StarsBtoTkin$ as a function of stellar mass with the black solid line showing the median relation. 
In addition, we define and calculate an ideal bulge-to-total stellar morphology, $\IdealStarsBtoTkin$, to represent the morphology a galaxy should ideally have if without influenced by the numerical heating effect.
For a given simulated galaxy, we record the initial circularity of all stellar particles at birth, $\epsilon_{\mathrm{circ,ini}}$.
Assuming that after birth, stellar particles maintain their initial kinematic state, the ideal morphology $\IdealStarsBtoTkin$ is defined as twice the mass fraction of stellar particles with $\epsilon_{\mathrm{circ,ini}} < 0$.
In panel (d), the median $\IdealStarsBtoTkin$ for galaxies with different masses are over-plotted by a blue dashed line.
To minimize the influence from external mergers or accretions, we only compute $\IdealStarsBtoTkin$ for galaxies with $\Mstars < 10^{10} \Msun$, as typically over $90\%$ of their stars form by internal star formation.

By comparing the blue line with the black one (i.e. the actual morphology simulated) in the panel (d) of Fig.~\ref{fig:size_vs_morph}, the impact of two-body scattering heating on galaxy morphology is seen as a function of galaxy stellar mass, with less massive galaxies affected by it more strongly, as shown in detail in panel (e).
As stellar mass increases, $\StarsSize / \Rconv$ increases, resulting in a smaller impact from two-body scattering.
Panels (d) and (e) also illustrate that the value of $\StarsBtoTkin$ is mostly contributed by $\IdealStarsBtoTkin$, more than the contribution from the differences between the two. 
Therefore, the overall trend that lower mass galaxies having more dispersion-dominated morphology is more determined by the intrinsic misaligned star formation, than by the numerical heating effect.

We then investigate the impact of resolution of simulation on our results, by applying similar analysis to simulations TNG50-2 and TNG50-3 (both having the same initial condition and box size as TNG50-1, but with lower resolution).
The comparison results are presented in Fig.~\ref{fig:compare_diff_simulations}.
Upper panels show that, galaxies with a given stellar mass in higher resolution simulations have a smaller convergence radius.
Moreover, the galaxy sizes in different simulations are generally converging towards each other, with only slight differences at the low-mass end \citep[detailed discussion of the size convergence between IllustrisTNG simulations can be found in][]{2018MNRAS.473.4077P, 2019MNRAS.490.3196P}.
Thus, galaxies in higher resolution simulations have larger values of $\StarsSize / \Rconv$, with stars less distributed within $\Rconv$ than their low-resolution counterparts.
With decreasing resolution from TNG50-1 to TNG50-3, the mass at which $\StarsSize / \Rconv$ goes above $1$ increases and roughly scales with particle mass, changing from $10^{8.8} \Msun$ in TNG50-1 to $10^{9.7} \Msun$ in TNG50-2 and $10^{10.7} \Msun$ in TNG50-3, corresponding to approximately $10^{4}$ stellar particles (see Appendix \ref{sec:Appendix_A} for details).
The morphology of galaxies more massive than these limits is relatively less affected by two-body scattering and thus more credible.

As shown in the bottom panels, simulated galaxies with lower resolution generally have larger $\StarsBtoTkin$ due to more numerical heating effect, while the $\IdealStarsBtoTkin$ in different simulations basically converge.
Fig.~\ref{fig:compare_diff_simulations} demonstrates that the kinematic morphology of simulated galaxies can be affected a lot by resolution, with galaxies being more dispersion-dominated for lower resolution.
For lower resolution simulation, the galaxy morphology that can only roughly be trusted for higher masses.
In addition, while numerical heating is severe for radius smaller than the convergence radius, the kinematic state of galaxy component in the inner region is not reflecting the real situation.
Therefore, when analysing the morphology of simulated galaxies, cautions must be made on the effect of resolution.

\section{Conclusions and Discussions}
\label{sec:Summary}

In this work, we study the kinematic morphology of low-mass central galaxies in the cosmological simulation TNG50-1, to investigate what drives the morphological differences in low-mass galaxies.
We select 595 central galaxies with $\Mstars = 10^{8.5} - 10^{9.0} \Msun$ at $z=0$ in TNG50-1. 
The majority of these simulated galaxies are dispersion-dominated, with only a minority displaying a rotation-dominated morphology, qualitatively consistent with observations.

By tracing and analysing the evolution of low-mass galaxies in detail, we find that these galaxies generally develop a cold gas disc early on, and most stars form within the rotating cold gas disc, regardless of the present morphology of galaxies.
However, the orientation of cold gas disc relative to the existing stellar component may evolve with cosmic time.
If the cold gas disc remains aligning with the galaxy during its evolution, stars formed at different times share the same rotational direction, leading to a rotation-dominated system.
On the contrary, in case of frequent misalignment between a galaxy and its cold gas disc, stars formed at different times have different rotation directions, resulting in a dispersion-dominated system.
Our results reveal a clear trend that low-mass galaxies with stronger misaligned star formation tend to be more dispersion-dominated (Fig.~\ref{fig:all_gals_misalignment_angle_vs_starsBtoTkin}).

Besides, we find that the two-body scattering effect has a non-negligible impact on the simulated morphology of low-mass galaxies.
This effect would numerically heat an originally disc-like component to become more dispersion-dominated, especially at small radii, within the convergence radius of a galaxy.
As a result, galaxy size has an additional effect on the simulated morphology of low-mass galaxies, with galaxies having larger size than the convergence radius less affected by the two-body scattering effect and thus retain a more rotation-dominated morphology (Fig.~\ref{fig:all_gals_misaligned_vs_hmr_vs_z0_stars_BtoTkin}).

The numerical effect that makes the simulated galaxies more dispersion-dominated than ideal cases affects more galaxies of lower masses, and galaxies simulated with lower resolutions.
Therefore, cautions must be made when analysing the morphology of simulated galaxies with low-mass and/or low-resolution. 
While galaxies with half-mass radius larger than the convergence radius can be considered less affected by the two-body scattering effect, our results suggest that the simulated galaxy morphology becomes roughly reliable when the number of stellar particles exceeds about $10^{4}$.

We should also note that, morphology of the central region of simulated galaxies such as galactic bulges and bars, can be severely affected by the numerical effect, and may not be trusted. 
For example, the convergence radius for numerous galaxies in TNG50-1 is greater than $1$ kpc.
The bulge component of galaxies, on the other hand, typically has an effective radius similar to or smaller than $1$ kpc \citep[e.g.][]{2010ApJ...716..942F, 2024MNRAS.529.4565H}.
Therefore, bulge components of galaxies can not be simulated robustly at this resolution level.
Simulations with higher resolutions are needed to study the inner bulge or bar components of galaxies.

\section*{Acknowledgements}


We thank the anonymous referee for the prompt and constructive report.
We thank Aaron D. Ludlow and Shi Shao for their helpful discussions and comments.
This work is supported by the National Natural Science Foundation of China (grant No. 11988101), the National SKA Program of China (Nos. 2022SKA0110200, 2022SKA0110201), the National Key Research and Development Program of China (No. 2023YFB3002500), the Strategic Priority Research Program of Chinese Academy of Sciences, (grant No. XDB0500203), and K.C. Wong Education Foundation.
The IllustrisTNG simulations were undertaken with compute time awarded by the Gauss Centre for Supercomputing (GCS) under GCS Large-Scale Projects GCS-ILLU and GCS-DWAR on the GCS share of the supercomputer Hazel Hen at the High Performance Computing Center Stuttgart (HLRS), as well as on the machines of the Max Planck Computing and Data Facility (MPCDF) in Garching, Germany.

\section*{Data Availability}


The IllustrisTNG simulations are publicly available and accessible at \url{https://www.tng-project.org/data/}.
The data produced in this work will be shared upon reasonable request to the corresponding author.



\bibliographystyle{mnras}
\bibliography{references}




\appendix

\section{Two-body relaxation and convergence radius}
\label{sec:Appendix_A}

The two-body scattering effect has been inspected by studying the two-body relaxation process quantitatively in detail, from theoretical point of view to simulations \citep[e.g.][]{1987gady.book.....B, 2008gady.book.....B, 2003MNRAS.338...14P, 2004MNRAS.348..977D,  2019MNRAS.487.1227Z, 2019MNRAS.488L.123L, 2019MNRAS.488.3663L}.

Theoretically, considering a test particle with a given velocity crossing a system of a given radius, the two-body relaxation time is used to describe the time required for velocity change induced by two-body scattering to become comparable to the initial velocity.
When the two-body relaxation time is sufficiently large, the system can be approximated as a collisionless system, meaning that the impact of two-body scattering effect is negligibly small.
Specifically, for an $N$-body system, \citet{2008gady.book.....B} derived the two-body relaxation time $t_{\mathrm{relax}} \propto N / \ln N$.

In hydrodynamical simulations, the simulated galaxies typically include dark matter particles and stellar particles, with the former being more massive than the latter \citep[see review by][]{2020NatRP...2...42V}.
For such system, \citet{2019MNRAS.488L.123L} obtained the two-body relaxation time $t_{\mathrm{relax}} \propto N_{\mathrm{dm}} / \ln N_{\mathrm{dm}}$, where $N_{\mathrm{dm}}$ is the number of dark matter particles, implying that the relaxation time is mainly determined by the more massive dark matter particles.
However, this result is based on an assumption that the total stellar mass in galaxies is much smaller than the total dark matter mass.
In more realistic cases, the stellar mass fraction may not be negligible, especially in the central region of galaxies \citep[e.g.][]{2018MNRAS.481.1950L, 2019MNRAS.484.1401R, 2024MNRAS.527..706Z, 2024MNRAS.528.5295Y}.
Therefore, in this work we consider also the role of stellar particles in the two-body relaxation process, as described in the following.

For a system with a radius of $r$, which includes $N_{\mathrm{dm}}$ dark matter particles of mass $m_{\mathrm{dm}}$ and $N_{\mathrm{stars}}$ stellar particles of mass $m_{\mathrm{stars}}$, if a test particle travels through this system with a velocity of $v$, the cumulative effect of two-body encounters will result in the following relative velocity change as derived by \citet[][]{2019MNRAS.488L.123L}:
\begin{equation}
\label{eq:rel_vel_change}
\frac{\Delta v^2}{v^2} = \frac{8}{N_{\mathrm{dm}}} \ln \Lambda \frac{(1 + \psi/\mu)}{(1 + \psi)^2}
\end{equation}
where the mass fraction in stars $\psi = N_{\mathrm{stars}} m_{\mathrm{stars}} / N_{\mathrm{dm}} m_{\mathrm{dm}}$, the particle mass ratio $\mu = m_{\mathrm{dm}} / m_{\mathrm{stars}}$, and the Coulomb logarithm $\Lambda = N_{\mathrm{dm}}(1 + \psi)/(1+{\mu}^{-1})$.
The initial conditions in IllustrisTNG simulations contain equal numbers of dark matter particles and gas cells, thus $\mu \approx \Omega_\mathrm{dm}/\Omega_\mathrm{baryon} = 5.3$ \citep[][]{2016A&A...594A..13P}.

Assuming the time required for each crossing is of the order of the orbital time $t_{\mathrm{orbit}} = 2 \pi r / \sqrt{GM(<r)/r}$, and the number of crossings needed for a significant velocity change is $1 / (\Delta v^2 / v^2)$, then the two-body relaxation time can be expressed as $t_{\mathrm{relax}} = t_{\mathrm{orbit}} / (\Delta v^2 / v^2)$.
Using the orbital time at virial radius $R_{200}$, which is comparable to the Hubble time, as the unit of two-body relaxation time, we have
\begin{equation}
\label{eq:relax_time}
\kappa_{\mathrm{relax}} = \frac{t_{\mathrm{relax}}}{t_{\mathrm{orbit}, 200}} = \frac{t_{\mathrm{orbit}} / (\Delta v^2 / v^2)}{t_{\mathrm{orbit}, 200}} = \sqrt{200} \left( \frac{\overline{\rho}{}}{{\rho}_{\mathrm{crit}}} \right) ^{-1/2} \left( \frac{\Delta v^2}{v^2} \right) ^{-1}
\end{equation}
where $\overline{\rho}$ is the mean mass density enclosed within $r$, and ${\rho}_{\mathrm{crit}}$ is the critical density of the Universe.
Based on equation (\ref{eq:relax_time}), the so-called convergence radius $r_{\mathrm{conv}}$ can be obtained by solving $\kappa_{\mathrm{relax}} = 1$ \citep[e.g.][]{2003MNRAS.338...14P, 2019MNRAS.487.1227Z, 2019MNRAS.488L.123L, 2019MNRAS.488.3663L}.

For simulated galaxies, we can use their convergence radius to measure the region affected by two-body scattering.
Beyond $r_{\mathrm{conv}}$, the two-body relaxation time would be similar to or even greater than the Hubble time, indicating a relatively small impact of two-body scattering.
While within $r_{\mathrm{conv}}$, especially near the center, the relaxation time can be quite short, meaning that two-body scattering may significantly affect the simulated kinematics.

\begin{figure}
\includegraphics[width=0.85\columnwidth]{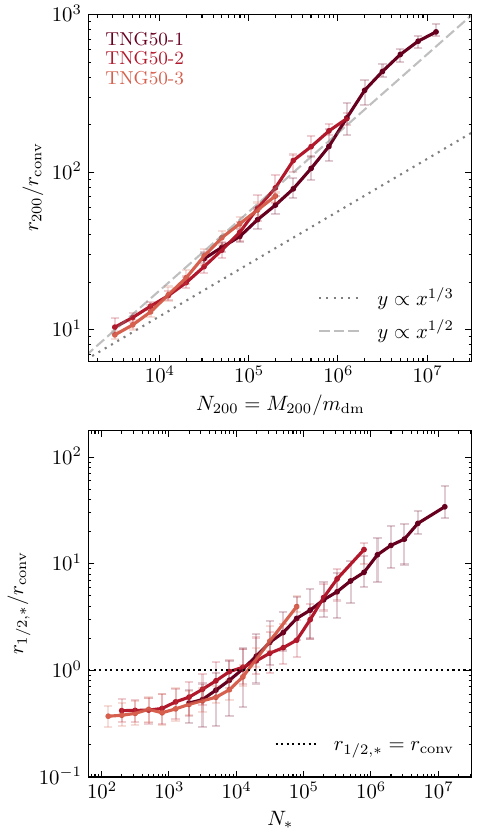}
\caption{
The dependence of $r_{200} / \Rconv$ on the particle number $N_{200} = M_{200} / m_{\rm{dm}}$ (top panel) and the dependence of $\StarsSize / \Rconv$ on the stellar particle number $N_*$ (bottom panel) in different IllustrisTNG50 simulations.
Dotted and dashed lines in the top panel represent power laws with exponents of $1/3$ and $1/2$, respectively, for reference.
}
\label{fig:diff_sim_z0_rconv_fitting}
\end{figure}

From the definition of convergence radius, we can derive that:
\begin{equation}
\frac{r_{200}}{\Rconv} = N_{200}^{1/3} N_{\rm{conv}}^{-1/3} \left( \frac{\Delta v^2}{v^2} \right) ^{-2/3}
\end{equation}
where $N_{200} = M_{200} / m_{\rm{dm}}$ and $N_{\rm{conv}} = M_{\rm{conv}} / m_{\rm{dm}}$.
Here, $M_{200}$ and $M_{\rm{conv}}$ represent the total mass enclosed within the virial radius $R_{200}$ and the convergence radius $\Rconv$, respectively.
If $N_{\rm{conv}}$ and $\Delta v^2 / v^2$ have no/weak dependence on $N_{200}$, then $r_{200} / \Rconv \propto N_{200}^{1/3}$, which is obtained also in other studies based on $N$-body simulations \citep[e.g.][]{2019MNRAS.488.3663L}.
However, for IllustrisTNG hydro-simulations, we note that the above relation can not be simply applied.
As shown in the top panel of Fig.~\ref{fig:diff_sim_z0_rconv_fitting}, results from TNG50-1, -2 and -3 exhibit that $r_{200} / \Rconv$ is roughly scaling of $N_{200}^{\alpha}$, but the exponent $\alpha$ is greater than $1/3$, and is actually close to $1/2$.
Additionally, these simulations show a similar dependence of $\StarsSize / \Rconv$ on the number of stellar particles $N_*$, as indicated in the bottom panel.
In general, when the number of stellar particles exceeds approximately $10^{4}$, the half-mass radius of galaxies exceeds the convergence radius.




\bsp	
\label{lastpage}
\end{document}